# Suppression of individual peaks in two-colour high harmonic generation

S Mitra[1,2,6], S Biswas[1,2], J Schötz[1,2], E Pisanty[3], B Förg[1,2], G A Kavuri[1], C Burger[1,2], W Okell[1,2], M Högner[1,2], I Pupeza[1], V Pervak[1,2,4], M Lewenstein[3,5], P Wnuk[1,2] and M F Kling[1,2,6]

[1] Max Planck Institute of Quantum Optics, Hans-Kopfermann-Str. 1, 85748 Garching, Germany
[2] Department of Physics, Ludwig-Maximilians-Universität Munich, Am Coulombwall 1, 85748 Garching, Germany
[3] ICFO–Institut de Ciencies Fotoniques, The Barcelona Institute of Science and Technology, Av. Carl Friedrich Gauss 3, 08860 Castelldefels (Barcelona), Spain
[4] Ultrafast Innovations GmbH, 85748 Garching, Germany
[5] ICREA, Passeig de Lluís Companys, 23, 08010 Barcelona, Spain

E-mail: sambit.mitra@mpq.mpg.de and matthias.kling@mpq.mpg.de

**Abstract**
This work investigates the suppression of individual harmonics, simultaneously affecting specific even and odd orders in the high-harmonic spectra generated by strongly tailored, two-colour, multi-cycle laser pulses in neon. The resulting spectra are systematically studied as a function of the electric-field shape in a symmetry-broken (ω–2ω) and symmetry-preserved (ω–3ω) configuration. The peak suppression is reproduced by macroscopic strong-field approximation calculations and is found to be unique to symmetry-broken fields (ω–2ω). Additionally, semi-classical calculations further corroborate the observation and reveal their underlying mechanism, where a nontrivial spectral interference between subsequent asymmetric half-cycles is found to be responsible for the suppression.

(Some figures may appear in colour only in the online journal)

## 1. Introduction

High harmonic generation (HHG) using intense laser pulses has long proven to be an extremely reliable source of high-energy, ultrashort and coherent light radiation ranging from UV to soft x-ray region [1–3]. As understood by the three-step model, HHG is a highly nonlinear process, which is strongly dependent on the electric field shape of the driving laser [4]. As a consequence, the most common approach to modify or control the temporal structure, spectral density and yield of HHG has been through altering the driving laser field. A prominent application of such control is in the production of isolated attosecond pulses by restricting the re-scattering process within a single half cycle using a gate in intensity [5] or polarization of the driving fields [6]. Other techniques involve mixing with an additional weak laser frequency component or colour, which perturbs the fundamental laser field to either modify or probe the highly nonlinear ionization probability, suppress electron re-scattering by an induced ellipticity of the laser field, or use phase matching properties unique to multi-coloured fields [7–11].

In contrast to numerous previous experimental works using the secondary colour as a weak perturbation, here, we explore a regime which involves significant modification of the driving electric field structure in a linearly polarized configuration. A strong deviation from the usual sinusoidal shape of the laser field leads to significantly altered ionization times, electron trajectories and their recollision kinetic energies. A few works carried out in a similar regime on two-colour control have reported various degrees of control over the HHG spectrum. The synthesis of a fundamental with its strong second harmonic component was shown to produce a strong cutoff energy oscillation [12, 13] and an overall shift of central energy of the HHG spectrum as a function of the waveform [14]. The overall shift in HHG spectral density was shown to be further enhanced under certain conditions, resembling a swallow-tail caustic within a parameter space of relative field strength, phase-delay of the two-colour driver, and photon energy [15]. Theoretical works have explored two-colour HHG numerically and analytically over a large range of parameters, and different parent atoms [16, 17]. Additionally, others have reported on the effect of unconventional non-gaussian temporal pulse envelopes [18] and the role of atomic Coulomb potential on the spectral density brought about by two-colour HHG [19]. Such control has already found applications in high-harmonic spectroscopy to probe multi-electron effects in noble gases [20, 21].

---

[6] Authors to whom any correspondence should be addressed.

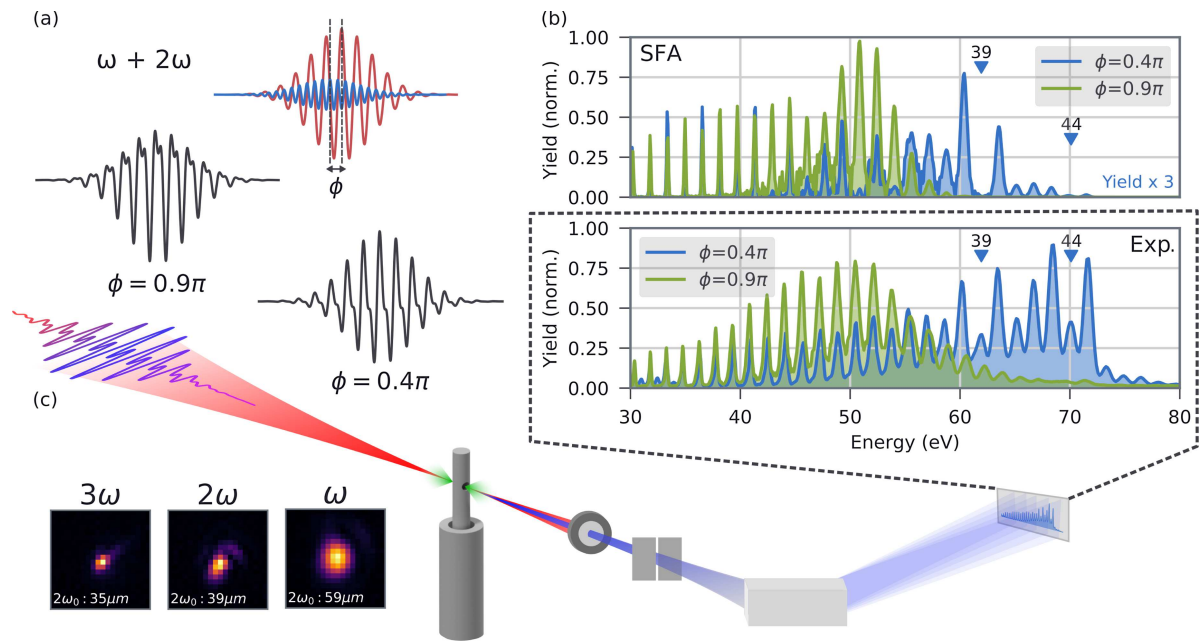

**Figure 1.** Phase-dependent single harmonic modulation in the HHG spectra generated by a two-colour driving field. (a) Illustration of waveforms resulting from an ω–2ω field at a phase-delay of 0.4π and 0.9π. (b) The respective HHG spectra obtained experimentally and from macroscopic SFA calculations. The spectrum obtained from simulations for $\phi = 0.4\pi$ has been magnified in yield by a factor of 3 for better visibility and comparison. The solid arrow markers indicate two examples of harmonic suppression occurring for an even and odd harmonic simultaneously. (c) A schematic overview of the experimental setup and beam profiles of each colour at the focus. The HHG spectra for ω–3ω driving fields are generated in a similar fashion.

The current report builds on the existing literature and investigates suppression of individual harmonics in HHG driven by an ω–2ω field, where simultaneous suppression of odd and even order harmonics are observed. Generally, only even harmonics are suppressed in a single colour driver, while none are suppressed in ω–2ω drivers as shown in the literature for weakly perturbing two-colour pulses [7, 22]. The behaviour of this unique suppression is studied as a function of the driving waveform controlled by the relative phase-delay. The resulting spectra are then reproduced with strong-field approximation (SFA) calculations including macroscopic and propagation effects. The degree of agreement between experiment and SFA calculations is quantitatively extracted using a fitting procedure. Furthermore, intuitive semi-classical calculations qualitatively reproduce the suppression, and hint at interference effects between harmonic emission from subsequent laser half-cycles. The experiment and calculations are repeated for an ω–3ω configuration, given its difference in subsequent half-cycle symmetry. Finally, a detailed semi-classical analysis of spectral modulations obtained in each half-cycle of ω–2ω and ω–3ω configurations are used to identify the mechanism responsible for the observed peak suppression in the former.

## 2. Methods

### 2.1. Experiment

The driving laser pulses used here were centered at 780 nm with a duration of 25 fs, generated by a commercial titanium sapphire laser (Femtopower Compact Pro HR, Spectra Physics) mixed with its second harmonic at 390 nm or third harmonic at 260 nm as illustrated for the former in figure 1(a). Waveform-control is achieved by delaying an arm in a three-colour, three-arm Mach–Zehnder interferometer. A detailed description of the multi-harmonic generator, interferometer and the laser source can be found in reference [23]. The setup used in this work has been modified to improve its interferometric stability by significantly reducing the path-length of each arm. The recorded spectra are continuously scanned over a delay of 4 fs around the peak intensity of the two laser pulses. The duration of the scan is restricted to five minutes to prevent significant averaging over slow drifts in the interferometer. The mirrors after the interferometer have been custom coated to maximize the reflection of all three colours. The maximum pulse energies obtained right before the experimental chamber were 265 $\mu$J ($\omega$), 65 $\mu$J ($2\omega$) and 35 $\mu$J ($3\omega$) with on-target focal spot sizes ($2w_0$) of 59 $\mu$m, 39 $\mu$m, and 35 $\mu$m, respectively. Unlike other studies involving non-collinear multi-colour interferometers using irises to reduce power, the interferometer arms use a combination of a half-wave plate and a polarizer (for $\omega$), and alignment of the respective BBO ($\beta$-barium borate) crystal (for $2\omega$ and $3\omega$) to control the power. This technique preserves the real-space overlap and respective focal spot sizes, leading to an accurate estimation of intensity (on target) and the relative field strength of the two colours. The maximum possible shift in spectrum introduced by the misaligned BBO to attenuate the intensity by a factor of ten, was calculated using the package Lab2 [24]. A maximum deviation of 10 nm and 5 nm is obtained for $2\omega$ and $3\omega$, respectively. Theoretical semi-classical calculations (as described in section 2.2) reveal negligible effects on the two colour HHG spectrum. Here, the target is a laser-drilled 1.5 mm thick stainless-steel gas cell filled with neon, followed by a thin metallic (aluminium) filter to block the driving laser pulses. The filtered XUV beam is then incident onto a commercial flat-field XUV spectrometer (McPherson Model 248/310) which records the spectrum (figure 1(b)) as a function of delay between two colours as shown in figure 1(c).

### 2.2. Calculation

The tailored pulses used in the experiments are defined as $E(t) = G_{env}(t) \cdot \frac{E_0}{1+R}(\cos(\omega t) + R\cos(N\omega + \phi))$, where $R = \frac{E_{N\omega}}{E_\omega}$ signifies the strength of a two-colour combination and thereby the degree of waveform modification. $N = 2$ and 3 for an $\omega$–$2\omega$ and $\omega$–$3\omega$ configuration respectively, $\phi$ is the phase-delay introduced by the interferometer between the two colours, $G_{env}$ is the temporal pulse envelope, and $E_0$ is the peak field strength of the tailored pulse.

The experiments are first reproduced theoretically with SFA simulations, where the time-dependent dipole response of an atom is calculated from an analytic description [25]. Saturation of the ground state is also taken into account [26]. Along with a single-atom response, macroscopic propagation effects in the generating medium are included in the simulation (considering cylindrical symmetry) for accurate reproduction of experimental conditions. The infrared driver ($P_{IR}$) and the generated XUV pulses ($P_{XUV}$) in the medium are propagated using a first-order propagation equation in co-moving coordinates at vacuum speed of light, having the form $\partial_z E_{IR/XUV} = -\frac{ic}{2\omega}\Delta_\perp E_{IR/XUV} - \frac{i\omega}{2c\epsilon_0}P_{IR/XUV}$, where $z$ is the propagation direction [27, 28]. Here, $P_{IR} = P_{IR,lin} + P_{Kerr} + P_{plasma}$. $P_{IR}$ includes linear effects, such as absorption and linear refraction, and nonlinear effects include spectral blue shifting, defocusing due to plasma formation [29] and lensing, and self-phase modulation due to the Kerr effect [30]. Similarly, $P_{XUV} = P_{XUV,lin} + P_{dipole}$, comprising of the linear response and a dipole component from SFA calculations. After traversing through the gas target, the XUV is numerically propagated to the far-field until it reaches the distance at which the spectrometer is placed (see figure 1(c)). A slight cropping of the XUV beam at the entrance of the spectrometer slit in the vertical direction in the experiment is also accounted for, as the two-colour phase-dependent ($\phi$) divergence of the XUV beam profile leads to a slightly modified spectral modulation with $\phi$ [31]. Finally, an additional transmission function for a partially oxidized aluminium filter is applied to the HHG spectrum for a more realistic calculation of spectral density, as used in the experiment [32]. Further details on macroscopic aspects of the simulation can be found in reference [27].

To intuitively understand the physics involved, semi-classical calculations are performed in addition, using the three-step model in a two-colour field [4]. The harmonic emissions from two subsequent half-cycles are separately calculated in the time domain as an analogy to the dipole response in SFA calculations [25, 33]. A Fourier transform of the temporal response provides the energy spectrum along with its spectral phase, which is essential for obtaining the harmonic peaks and other modulations in spectra from the coherent addition of emission from multiple half-cycles. The simplified temporal response for each half-cycle burst can be expressed as: $d(t_R) = \sqrt{W(t_B)\frac{dt_B}{dt_R}}e^{j\frac{1}{\hbar}S(t_B,t_R)}$, where $t_B$ and $t_R$ are the electron birth- and recollision-times respectively, obtained by classically propagating the electron trajectories in the laser field. $S(t_B, t_R) = \int_{t_B}^{t_R} E_{KE}(t)\,dt + I_P(t_R - t_B)$, is the classically acquired phase by the electron through its excursion and eventual recollision, $W(t_r)$ is the Ammosov–Delone–Krainov (ADK) tunnelling rate [34, 35], $I_P$ is the ionization potential of the parent atom, and $E_{KE}(t)$ is the instantaneous kinetic energy of the electron in the two-colour laser field. Such a semi-classical approach tends to predominantly favour long-trajectories over short, in contrast to experimental conditions where long-trajectories suffer from intensity averaged dephasing leading to inaccurate estimation of the correct HHG rate [36, 37]. However, they are still very useful for qualitatively probing the contribution of each half-cycle to the HHG spectra [12–14].

## 3. Results and discussion

### 3.1. Harmonic suppression in symmetry-broken fields ($\omega$–$2\omega$)

The suppression of individual harmonics in the two-colour ($\omega$–$2\omega$) HHG spectra beyond 60 eV, as shown in figure 1(b), is the focus of this work. Typically, the spectrum from a single colour laser pulse should exhibit a plateau of harmonic combs followed by an exponential drop-off in yield near the cut-off [5]. The rising yield of HHG with energy observed in figure 1(b), is characteristic of the XUV transmission function of a thin aluminium filter, followed by a sharp cutoff near its absorption edge around 73 eV. The other prominent feature is the difference in spectral density of two complementary waveforms where $\phi = 0.4\pi$ and $0.9\pi$. The variation of spectral density with $\phi$ simultaneously over multiple harmonics, causing a forked-like structure near the cutoff, has been observed in earlier works [12, 14, 15]. It has been attributed to the modification of recolliding electron trajectories and their kinetic energies, which can also lead to significant enhancement in the spectrum well below the classical cutoff, resembling swallowtail caustics for $R$ larger than 0.44 [15]. The overall change of spectral density in two-colour HHG has also been discussed in reference [14], while a strong modulation of the harmonic cutoff energies has been shown in reference [12]. The suppression of individual harmonics, however, remained unexplored in previous experimental works, and is a novelty of this report.

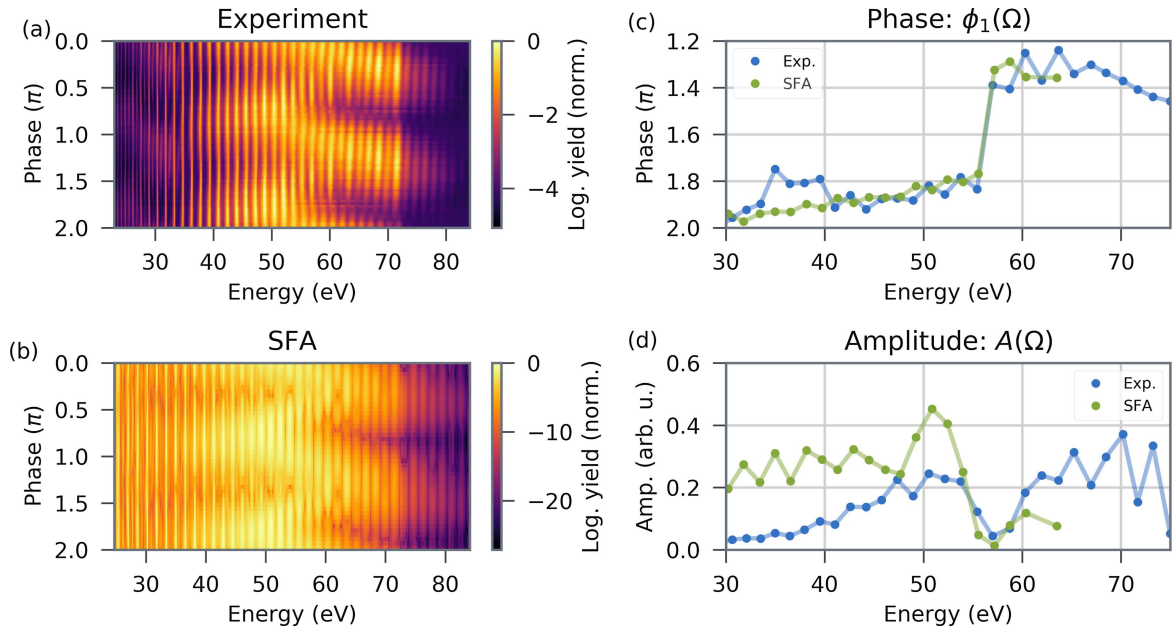

**Figure 2.** Modulation of HHG spectra as a function of relative phase-delay in symmetry-broken two-colour laser fields ($\omega$–$2\omega$). (a) Experimentally measured HHG spectra as a function of phase-delay and XUV energy. (b) Macroscopic SFA calculations for $R = 0.23$ and $I = 3 \cdot 10^{14}$ Wcm$^{-2}$. The data in (a) and (b) are both normalized to their respective global maxima. (c) Comparison of harmonic-dependent HHG yield-modulation-phase (parameter $\phi_1(\Omega)$ from fit function as described in the text) in the experimental and calculated HHG spectra. (d) Comparison of the corresponding amplitude (parameter $A(\Omega)$), as generated by the fitting procedure from the normalized values in (a) and (b).

The choice of $\phi = 0.4\pi$ in figure 1 corresponds to a waveform where the harmonic emission rate is very similar in the subsequent half-cycles, also leading to the strongest peak suppression behaviour. The complementary configuration of $\phi = 0.9\pi$ corresponds to a waveform with highly suppressed harmonic emission rate from alternate half-cycles, also corresponding to the experimental spectrum with the least peak suppression as illustrated in figure 1(b). Normally, one would intuitively expect these conditions to occur at $\phi = 0.5\pi$ (and $\pi$), where the ionization probability (ADK) is indeed equal (and alternating) in magnitude for the subsequent half-cycles, however, the ionization probability density also shifts in time. This leads to electrons with the same recollision kinetic energy in the subsequent half-cycles to experience different regions of the probability density function, resulting in unequal (and not fully alternating) electron recollision rates, as shown theoretically in reference [17].

The behaviour of harmonic suppression is further investigated by recording the experimental HHG spectra over a set of phase-delays ($0 < \phi < 2\pi$), which covers all the unique waveforms of an $\omega$–$2\omega$ field for fixed $R$. The resulting spectra are all shown together in a false-colour plot (figure 2(a)). The continuous scan over $\phi$ in figure 2(a) also shows the transition between $0.4\pi$ and $0.9\pi$, where a gradual increase or decrease in contrast of certain harmonics is visible beyond 60 eV. The supplementary information in reference [15] shows phase plots obtained under similar experimental conditions but their limited resolution makes it hard to infer such features. The theoretically predicted two-colour HHG rates in reference [17] strongly resemble the data presented here. A strong nontrivial interference pattern leading to suppression of individual harmonics is also visible, however, the mechanisms responsible are not discussed in detail. Our experimental results are further reproduced using macroscopic SFA calculations as described in section 2.2 using similar experimental parameters and shown in figure 2(b) for $R = 0.23$. The phase ($\phi$) axis in the experimental plot has been readjusted to match the simulation due to the absence of any absolute waveform detection in the experiment. A clear modulation in single harmonics also appears here and at the same energies and relative phase ($\phi$) as observed in the experiment.

The degree of correlation between the experiment (figure 2(a)) and macroscopic SFA calculations (figure 2(b)) is quantitatively determined using a fitting procedure. The modulation in yield for each individual harmonic as a function of phase-delay ($\phi$) between the colours is fit to a Fourier series of the following form:

$$f_{\text{fit}}(\phi, \Omega) = A(\Omega)\cos(\phi + \phi_1(\Omega)) + B(\Omega)\cos(2\phi + \phi_2(\Omega)) + C(\Omega). \quad (1)$$

The function captures the complete $\phi$ and $\Omega$ (harmonic frequency) dependent HHG spectra and converges well with negligible error. The first term maps the single prominent modulation below 55 eV, and then continuing further after a phase jump for higher energies. The second term, with twice the periodicity resolves the prominent forked structure, which appears beyond 55 eV as seen in figures 2(a) and (b). Although it is necessary to use the two terms for a converging fit, the coefficients of the first term are found enough for a quantitative comparison of experiment and theory. $A(\Omega)$ signifies the magnitude of the fundamental modulation component as extracted from the normalized spectra in figures 2(a) and (b). Similarly, $B(\Omega)$ indicates the magnitude of the second oscillation component which plays a role in fitting the modulations around 55 eV, while $C(\Omega)$ is a DC offset to account for the non-modulating part of the signal. The coefficients $B(\Omega)$ and $C(\Omega)$ are non-zero and obtain approximate maximum values of 0.1 and 0.5, respectively, at the harmonics around 54 eV. Although essential for the fit, one of the amplitude coefficients is found to be enough for a quantitative correlation between the experiment and SFA calculations. The phase offset terms $\phi_1(\Omega)$ and $\phi_2(\Omega)$ signify the phase drift of the respective oscillation components. In particular, $\phi_1(\Omega)$ is found to be a useful parameter, which accurately maps the gradual phase shift of the fundamental oscillation component in the experiment and calculated data (see figure 2(c)). A small change in the peak-intensity or $R$ of the calculated HHG spectra results in large observable shifts in $\phi_1(\Omega)$, serving as a sensitive correlation marker.

A direct comparison with their coefficients in figures 2(c) and (d) quantitatively exhibit good correlation. As compared to $\phi_1(\Omega)$, a larger deviation is observed in $A(\Omega)$. Given the complex balance of phase-matching parameters playing a role in macroscopic build-up of the HHG in the gas medium, small deviations in the theoretical modelling could lead to large differences in the calculated spectral density [36]. The slight difference observed in $\phi_1(\Omega)$ near 35 eV is due to the second-order diffraction of the same spectrum from the spectrometer grating overlapping onto the first. A peak detection algorithm is used to restrict the fitting only to the harmonics for correct unwrapping of the phases ($\phi_1(\Omega)$) extracted from each. Beyond 65 eV, the peak detection scheme for the simulated spectra fails to obtain a fit given the low signal in this region. The increasing deviation in yield (also evident in $A(\Omega)$) between the experiment and calculations near the cutoff energy remains an open question. The exact pressure at the gas target is hard to determine accurately in the experiment, which could result in inaccurate phase-matching conditions in the simulations for a certain spectral region.

The good quantitative correlation between experiment and theory lets us accurately fix critical parameters like intensity, relative field strength of the two colour configuration ($R$), and absolute phase $\phi$. These parameters are further used in semi-classical calculations to intuitively understand the mechanisms responsible for the rich interference features observed at certain values of $\phi$ (see figure 3). They also provide an insight into the general structure of the pattern intuitively, where re-colliding electrons form two distinct plateaus and two cutoff structures arising from two subsequent half-cycles of the laser field as shown in figures 3(a) and (b) as a fork shaped structure. However, close to $\phi = 0.4\pi$ and $1.4\pi$, the additional modulation pattern disappears in figure 3(b), which shows calculations for only emission from negative half-cycles. This hints at an inter-half-cycle interference at play. The time-energy structure of the classical electron recollisions for two complementary phase-delays ($\phi$) are shown in figures 3(c) and (d), which in case of $\phi = 0.4$ highlights the comparable ionization rate of harmonic emission (recolliding electrons) from subsequent half-cycles. However, the time-energy structure of the recolliding electrons, including the cutoff, is significantly different. In comparison, at $\phi = 0.9$, the harmonic emission is largely restricted to a single half-cycle.

### 3.2. Harmonic suppression in symmetry-preserving fields ($\omega$–$3\omega$)

To further investigate the role of interference from two subsequent half-cycles, the experiment and calculations were repeated for symmetry-preserving ($\omega$–$3\omega$) driving fields for similar intensity and $R$ values. The resulting spectra, shown in figure 4(a), exhibit far less features as compared to the $\omega$–$2\omega$ case. The macroscopic SFA calculations shown in figure 4(b) reveal a similar pattern. The reduced contrast between figures 4(a) and (b) can be attributed to larger interferometer drifts between $0 < \phi < 2\pi$ for smaller wavelengths. As discussed earlier, the fitting procedure is also employed here to judge the degree of correlation between experiment and theory (see figure 4(c)). The phase coefficients ($\phi_1(\Omega)$) are well reproduced beyond 50 eV. In contrast to $\omega$–$2\omega$ fields, here, the recolliding electron trajectories and their respective ionization rates vary identically with $\phi$ in the subsequent half-cycles. It leads to a strong suppression of HHG over large values of $\phi$ (in between $\pi$ and $\pi/2$ as seen in figures 4(a), (b) and (d)). In case of other values of $\phi$, the identical contributions constructively add up having a higher ionization rate without any visible individual harmonic suppression. The semi-classical calculation in figure 4(d), however, shows strong periodic modulation and harmonic suppression between 40 eV and 60 eV around $\phi = 1.5\pi$, which are absent in the experimental and SFA spectra. This modulation resembles intra-cycle, long-short trajectory interference as shown later quantitatively. The intra-cycle interference features observed for $\omega$–$2\omega$ fields in figures 3(a) and (b)) are also absent in the respective experimental and SFA spectra in figure 2 for $\omega$–$2\omega$ fields. This can be attributed to stronger dephasing of long-trajectories as compared to the short ones when averaged over varying intensities in the focal volume [36, 37]. The absence of additional intricate interference features in the semi-classical spectra from symmetric fields in figure 4(d), as compared to the ones observed at $\phi = 0.4\pi, 1.4\pi$ from asymmetric fields, further hints at inter-half-cycle interference at play.

### 3.3. Semi-classical interpretation of harmonic suppression

Apart from the realistic harmonic spectra shown in figures 3 and 4(d), the semi-classical calculations provide deeper insight into the source of harmonic peak suppression. The modulations in harmonic spectra are investigated by selectively choosing the spectrum generated by each half-cycle followed by their coherent addition. In figure 5(a), the different half-cycle spectra from an $\omega$–$2\omega$ combination are shown for $\phi = 0.4\pi$. This $\phi$ corresponds to the spectrum where strong harmonic peak suppression is observed in the experiment and macroscopic SFA calculations (see figure 1(b)). The subsequent half-cycles of the resulting $\omega$–$2\omega$ waveform exhibit similar electron recombination rates, as evident from the height of the curves labelled $T/2^+$ and $T/2^-$, where $T$ corresponds to one laser cycle of the IR ($\omega$). The modulations in the curves arise from long-short trajectory interference within each respective half-cycle. A difference in periodicity is observed in $T/2^+$ as compared to $T/2^-$ arising from different electric field shapes (see figure 3), which result in altered energy and phase of the recombining electron trajectories in the subsequent half-cycles (see figure 5(a)). The spectrum $T$ is obtained from a coherent sum of $T/2^+$ and $T/2^-$ (with a significant spectral phase difference), the subsequent half-cycles, resulting in strong spectral modulation and suppression of odd and even order harmonics. Finally, the realistic photon-energy spaced harmonic spectrum $3T$ is produced by the coherent addition of spectrum $T$ over three laser cycles. Looking back at spectrum $T$ again, we notice that this curve defines an envelope governing the suppression of harmonic order in spectrum $3T$. This appears from the spectral phase difference ($\Delta\varphi(\Omega)$) of $T/2^+$ and $T/2^-$ (also shown in figure 5(a)), which leads to a suppression of odd harmonics when $\Delta\varphi(\Omega) = 0$ and suppression of even harmonics when $\Delta\varphi(\Omega) = \pi$. These conditions can also be derived from the frequency comb expression of the spectra in question. The envelope or low frequency modulation of spectrum $T$ arises from the amplitude modulations in $T/2^+$ and $T/2^-$. This highlights the role of intra-half-cycle and inter-half-cycle interference as the source of strong individual harmonic suppression. However, as discussed earlier, the experiments and realistic SFA calculations, involving intensity averaging over the laser focal volume suppress long-trajectory contributions [36, 37]. This reduces the contribution of intra-half-cycle interference. Thus, inter-half-cycle interference is considered responsible for the prominent peak suppression and seemingly aperiodic modulation near $\phi = 0.4\pi, 1.4\pi$ for an $\omega$–$2\omega$ driver. The spectra for all values of $\phi$ (see figure 3) are in good qualitative agreement with experimental observations. Given the simplicity of our semi-classical calculations, a quantitative agreement with the experiment is not expected. This is primarily due to a large expected deviation in the proportion of long and short trajectories within each half-cycle.

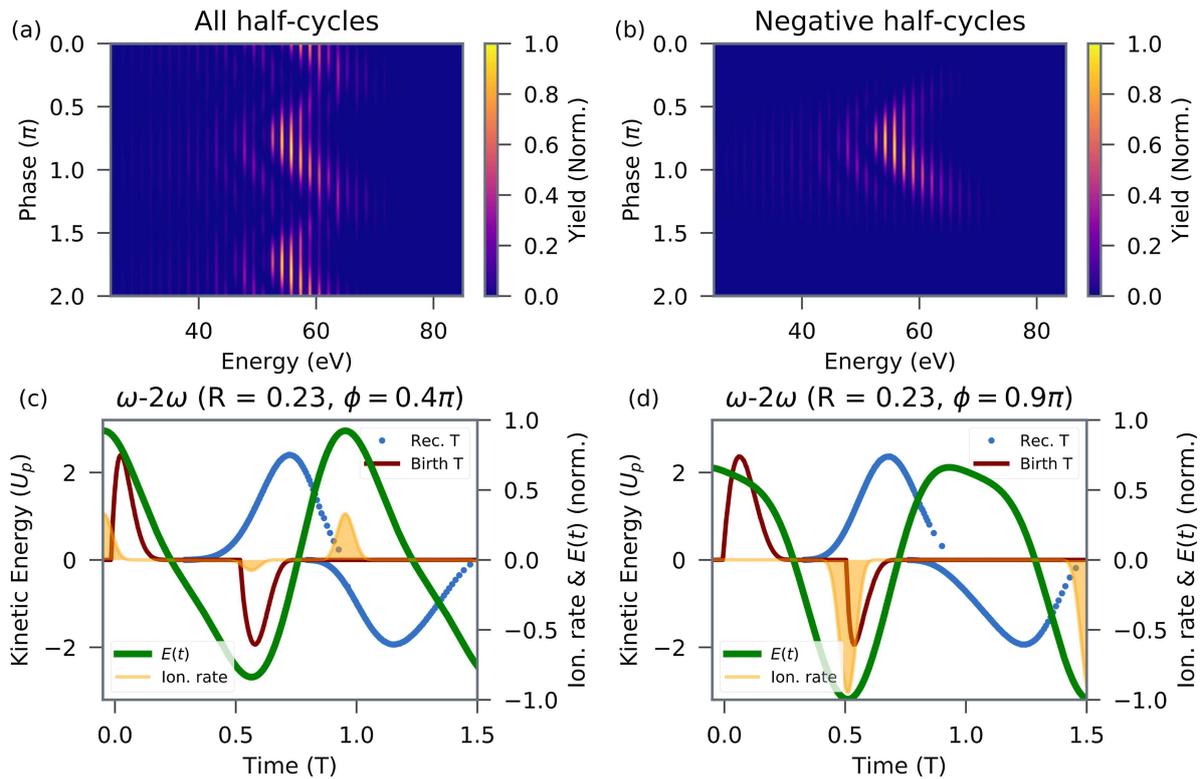

**Figure 3.** Semi-classical HHG rate for two-colour symmetry-breaking fields ($\omega$–$2\omega$). (a) Harmonic emission calculated from classically recolliding electrons as a function of $\phi$ from two subsequent half-cycles, coherently added over three complete laser cycles. (b) Harmonic emission exclusively from negative half-cycles, coherently added over three complete laser cycles. (c), (d) Analytically calculated time-dependent electron recollision energies as a function of their time of birth (Birth T) and their time of recollision (Rec. T) in 1.5 laser cycles. Also shown are the respective ionization rate (ADK) and electric field shape for a phase-delay of $\phi = 0.4\pi$ (c) and $\phi = 0.9\pi$ (d). The energy axis has been normalized to the ponderomotive energy ($U_p$), while the ionization rate has been normalized to the highest among all values of $\phi$. The negative values on the y-axis have been used only to decouple the emission directions.

Similarly, figure 5(b) illustrates the semi-classically calculated half-cycle spectra, driven by $\omega$–$3\omega$ at $\phi = 1.5$. This particular value of $\phi$ has been chosen as it corresponds to the appearance of strong modulation in the harmonic spectrum as observed in figure 4(d). Given the symmetry of subsequent half-cycles, the spectra $T/2^+$ and $T/2^-$ have exactly identical yield and intra-cycle interference pattern arising from long and short trajectory contributions. The coherent addition of $T/2^+$ and $T/2^-$ results in spectrum $T$, which exhibits an envelope with the same periodicity as observed from each half-cycle.

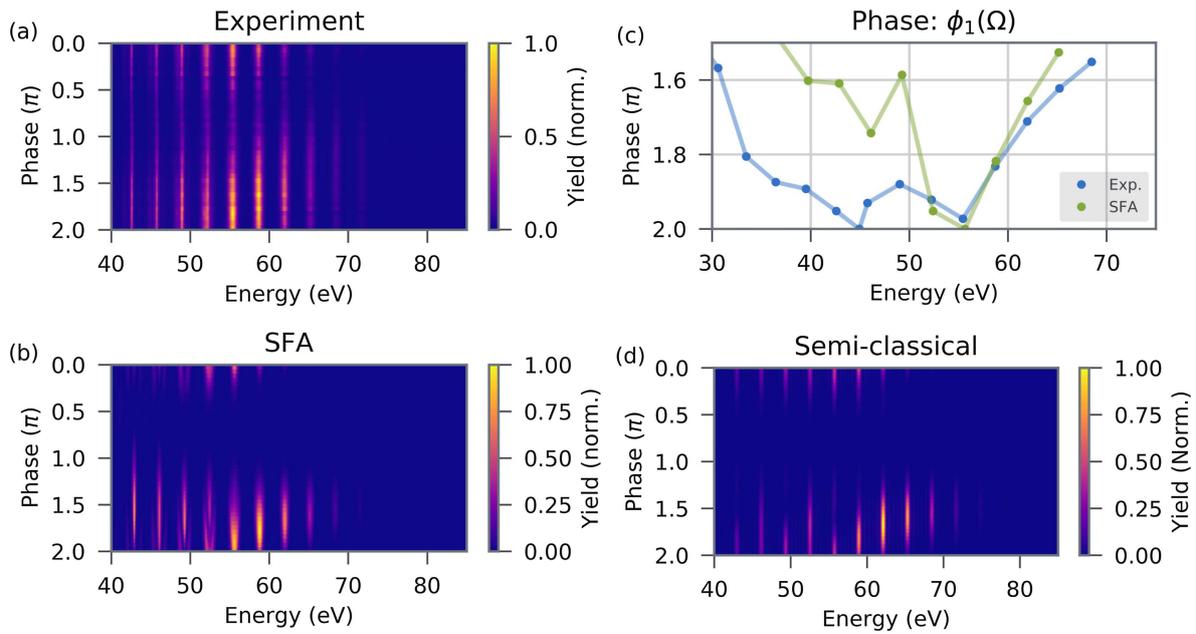

**Figure 4.** Phase-dependent HHG spectra in symmetry-preserving ω–3ω driving fields. (a) Experimentally obtained spectra. (b) Macroscopic SFA calculations for $R = 0.23$ and $I = 3 \cdot 10^{14}$ Wcm$^{-2}$. (c) Comparison of $\phi_1(\Omega)$ between experiment and macroscopic SFA calculations. (d) Semi-classical calculations including harmonic emission from two subsequent half-cycles coherently added over three complete laser cycles. Unlike in figure 2, the HHG yields in (a) and (b) are plotted on a linear scale as they produce better visibility of phase-dependent features.

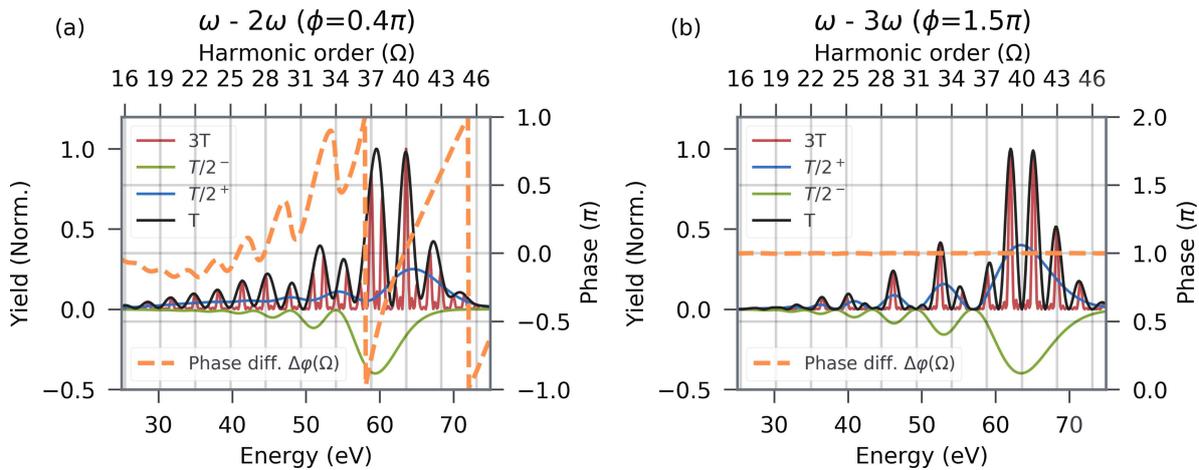

**Figure 5.** Semi-classical analysis of subsequent half-cycle contributions to spectral modulation in two-colour HHG. The spectra and spectral phase differences are obtained for (a) asymmetric half-cycles (ω–2ω), where $\phi = 0.4\pi$ and (b) symmetric half-cycles (ω–3ω), where $\phi = 1.5\pi$. The calculations here use the same intensity and $R$ values as in figures (3) and (4). The spectra shown here arise from coherent addition of all recombining electron trajectories within the positive half-cycle ($T/2^+$), negative half-cycle ($T/2^-$), one cycle ($T$), and three cycles ($3T$) where $T$ represents a cycle of the fundamental field ($\omega$). $3T$ and $T$ are normalized to their own respective maxima, while the other two are normalized to $T/2^-$ (and scaled by a factor 0.5) to maintain their relative yield. The spectral phase difference ($\Delta\varphi(\Omega)$) is calculated between $T/2^-$ and $T/2^+$. A negative yield is used to decouple the emission directions.

Given the identical energy and constant spectral phase difference ($\Delta\varphi(\Omega) = \pi$) of recombining electron trajectories in the two half-cycles, their coherent addition only leads to suppression of even harmonics, unlike in ω–2ω fields. The realistic sharp harmonic peaks are produced by the coherent addition of three full cycles and represented by $3T$ (in red). The low frequency modulation of $T$ arises from the modulating amplitudes of $T/2^+$ and $T/2^-$. Thus, intra-half-cycle interference is responsible for the modulation observed in figure 4(d) at $\phi = 1.5\pi$. As a consequence, the absence of this modulation in the intensity (in the focal volume) averaged experiment and realistic SFA spectra are expected as discussed earlier. The disappearance of the intra-half-cycle modulation is also observed in figure 4(d), where $\phi$ is close to 0 or $2\pi$. This occurs due to a strong suppression of short, and enhancement of long trajectories by the window of ionization-rate (ADK) in time, which shifts to earlier electron birth-times. Thus, a symmetric two-colour field (as shown here for $\omega$–$3\omega$ fields) is unable to support the nontrivial interference pattern obtained in a symmetry-broken field (as shown for $\omega$–$2\omega$ fields).

## 4. Conclusion

We observed the suppression of individual harmonics of both orders (odd and even) in the HHG spectra generated by two-colour symmetry-broken ($\omega$–$2\omega$) laser fields in neon, and demonstrated precise control over this peak-suppression contrast through a phase-delay ($\phi$) between the driving colours. The complete phase ($\phi$) dependent behaviour was well reproduced using macroscopic SFA calculations including propagation effects, and an intuitive semi-classical analysis of emission bursts from each half-cycle. The investigations revealed a nontrivial interference pattern to be responsible for the peak suppression at certain energies, which arise from strong spectral phase mismatch between bursts from subsequent half-cycles for symmetry-broken laser fields ($\omega$–$2\omega$). The symmetry requirement was further verified using symmetry-preserving ($\omega$–$3\omega$) fields where the interference pattern leading to harmonic peak suppression was absent.

This study highlights the role of symmetry, and inter-half-cycle interference in two-colour HHG, where the ratio of their central wavelengths turns out to be an important parameter in addition to intensity, $R$, and $\phi$. Other schemes using non commensurate combination in this regime would be an interesting extension to this study. Additionally, a stronger value of $R$ leads to the coalescence of more that two trajectories within a single half-cycle [15] and their subsequent interference with neighbouring half-cycles is another interesting exploration direction. Given its contribution to the harmonic phase, a change in the parent atom could also lead to largely modified interference envelope as already seen in theoretical predictions [17].

The demonstrated additional control parameter of half-cycle asymmetry will eventually be useful in optimising waveforms to generate a desired spectral modulation as have been performed for optimizing yield and cutoff energies [13, 38, 39]. It is expected to greatly benefit applications demanding further tunability of HHG, like in molecular spectroscopy or measurement of atomic transition lines with XUV frequency combs [40, 41].

## Acknowledgments

We acknowledge support from the German Research Foundation (DFG) via LMUexcellent and SPP1840. S M, J S, B F, and M F K acknowledge support from the Max Planck Society. S B is grateful for support by the Alexander von Humboldt foundation and MULTIPLY fellowship programme under Marie Skłodowska-Curie COFUND. P W acknowledges a Mobility Plus fellowship (no. 1283/MOB/IV/2015/0) from the Polish Ministry of Science and Higher Education. E P and M L acknowledge support of The Spanish Ministry MINECO (National Plan 15 Grant: FISICATEAMO No. FIS2016-79508-P, FPI), European Social Fund, Fundació Cellex, Generalitat de Catalunya (AGAUR Grant No. 2017 SGR 1341 and CERCA/Programme), ERC AdG OSYRIS and NOQIA, and the National Science Centre, Poland-Symfonia Grant No. 2016/20/W/ST4/00314. E P acknowledges Cellex-ICFO-MPQ Fellowship funding.

## ORCID iDs

S Mitra https://orcid.org/0000-0003-0837-2661
S Biswas https://orcid.org/0000-0002-7176-3887
E Pisanty https://orcid.org/0000-0003-0598-8524
M Högner https://orcid.org/0000-0002-6243-802X
M F Kling https://orcid.org/0000-0002-1710-0775

## References

[1] McPherson A, Gibson G, Jara H, Johann U, Luk T S, McIntyre I A, Boyer K and Rhodes C K 1987 *J. Opt. Soc. Am.* B **4** 595–601
[2] Li X F, L'Huillier A, Ferray M, Lompré L A and Mainfray G 1989 *Phys. Rev.* A **39** 5751–61
[3] Popmintchev T et al 2012 *Science* **336** 1287–91
[4] Corkum P B, Burnett N H and Ivanov M Y 1994 *Opt. Lett.* **19** 1870
[5] Krausz F and Ivanov M 2009 *Rev. Mod. Phys.* **81** 163–234
[6] Gilbertson S, Wu Y, Khan S D, Chini M, Zhao K, Feng X and Chang Z 2010 *Phys. Rev.* A **81** 043810
[7] Dudovich N, Smirnova O, Levesque J, Mairesse Y, Ivanov M Y, Villeneuve D M and Corkum P B 2006 *Nat. Phys.* **2** 781–6
[8] Kondo K, Kobayashi Y, Sagisaka A, Nabekawa Y and Watanabe S 1996 *J. Opt. Soc. Am.* B **13** 424–9
[9] Brizuela F et al 2013 *Sci. Rep.* **3** 01410
[10] Wei P, Miao J, Zeng Z, Li C, Ge X, Li R and Xu Z 2013 *Phys. Rev. Lett.* **110** 233903
[11] Diskin T and Cohen O 2014 *Opt. Express* **22** 7145–53
[12] Mauritsson J, Johnsson P, Gustafsson E, L'Huillier A, Schafer K J and Gaarde M B 2006 *Phys. Rev. Lett.* **97** 013001
[13] Ishii N, Kosuge A, Hayashi T, Kanai T, Itatani J, Adachi S and Watanabe S 2008 *Opt. Express* **16** 20876–83
[14] Mansten E, Dahlström J M, Johnsson P, Swoboda M, L'Huillier A and Mauritsson J 2008 *New J. Phys.* **10** 083041
[15] Raz O, Pedatzur O, Bruner B D and Dudovich N 2012 *Nat. Photon.* **6** 170–3
[16] Figueira de Morisson Faria C, Milošević D B and Paulus G G 2000 *Phys. Rev.* A **61** 063415
[17] Frolov M V, Manakov N L, Silaev A A and Vvedenskii N V 2010 *Phys. Rev.* A **81** 063407
[18] Hamilton K R, van der Hart H W and Brown A C 2017 *Phys. Rev.* A **95** 013408
[19] Birulia V A and Strelkov V V 2019 *Phys. Rev.* A **99** 043413
[20] Faccialà D et al 2016 *Phys. Rev. Lett.* **117** 093902